\documentstyle[twocolumn,floats,aps]{revtex}

\begin{document}
\draft

\twocolumn[\hsize\textwidth%
\columnwidth\hsize\csname@twocolumnfalse\endcsname

\title{\bf Noise Reduction and Universality in Limited Mobility 
Models of Nonequilibrium Growth}

\author{P. Punyindu and S. Das Sarma}
\address{Department of Physics, University of Maryland, College Park,
Maryland 20742-4111}

\date{\today}
\maketitle

\begin{abstract}
We show that a multiple hit 
noise reduction technique involving the acceptance of
only a fraction of the allowed atomistic deposition events could,
by significantly suppressing the formation of high steps and 
deep grooves,
greatly facilitate the identification of the universality class of
limited mobility discrete solid-on-solid 
conserved nonequilibrium models of
epitaxial growth. In particular, the critical growth exponents of the
discrete one dimensional molecular beam epitaxy growth model are
definitively determined using the noise reduction technique, and the
universality class is established to be that of the nonlinear
continuum fourth order conserved epitaxial growth equation.
\end{abstract}
\pacs{PACS: 05.40.+j, 81.10.Aj, 81.15.Hi, 68.55.-a}
\vskip1pc]
\narrowtext

In 1991 Das Sarma and Tamborenea introduced \cite{1} an extremely
simple one dimensional (d=1+1) instantaneous relaxation limited mobility
conserved
discrete solid-on-solid model of ideal molecular beam epitaxial (MBE)
growth under random vapor deposition nonequilibrium growth conditions.
In spite of the deceptively simple deposition and
relaxation/incorporation rules controlling its growth dynamics
the universality class of this discrete growth model, particularly in 
d=1+1 dimensions, has remained controversial \cite{2,3} and unresolved
in spite of a substantial body of work 
\cite{4,5,6,7,8,9,10,11}. Our lack of understanding of the
universality class of this one dimensional growth model \cite{1} is
particularly mysterious for the following three reasons: 
\begin{figure}

 \vbox to 5.5cm {\vss\hbox to 6cm
 {\hss\
   {\includegraphics{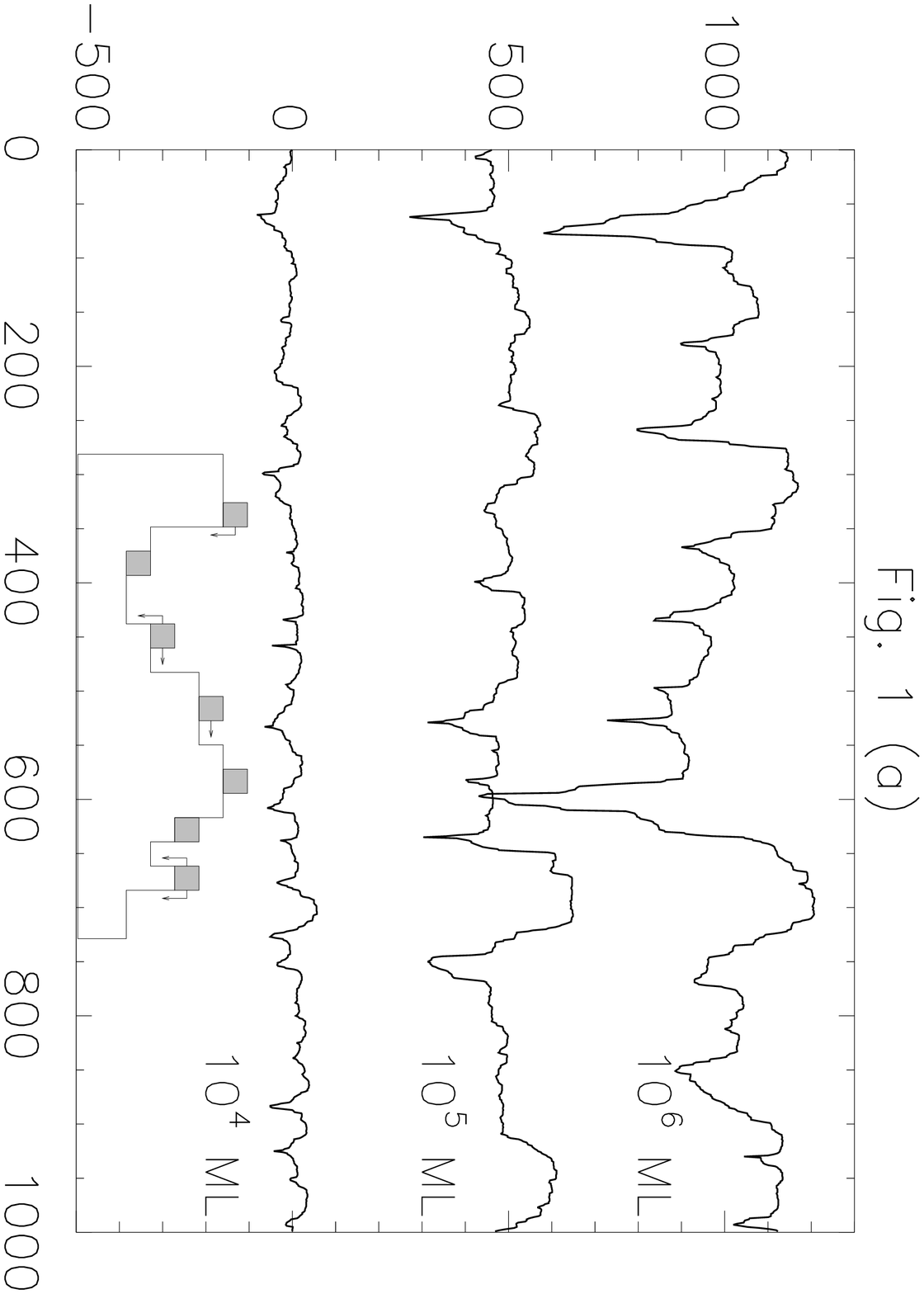}
   }
  \hss}
 }

\vskip .5cm

 \vbox to 5.5cm {\vss\hbox to 6cm
 {\hss\
   {\includegraphics{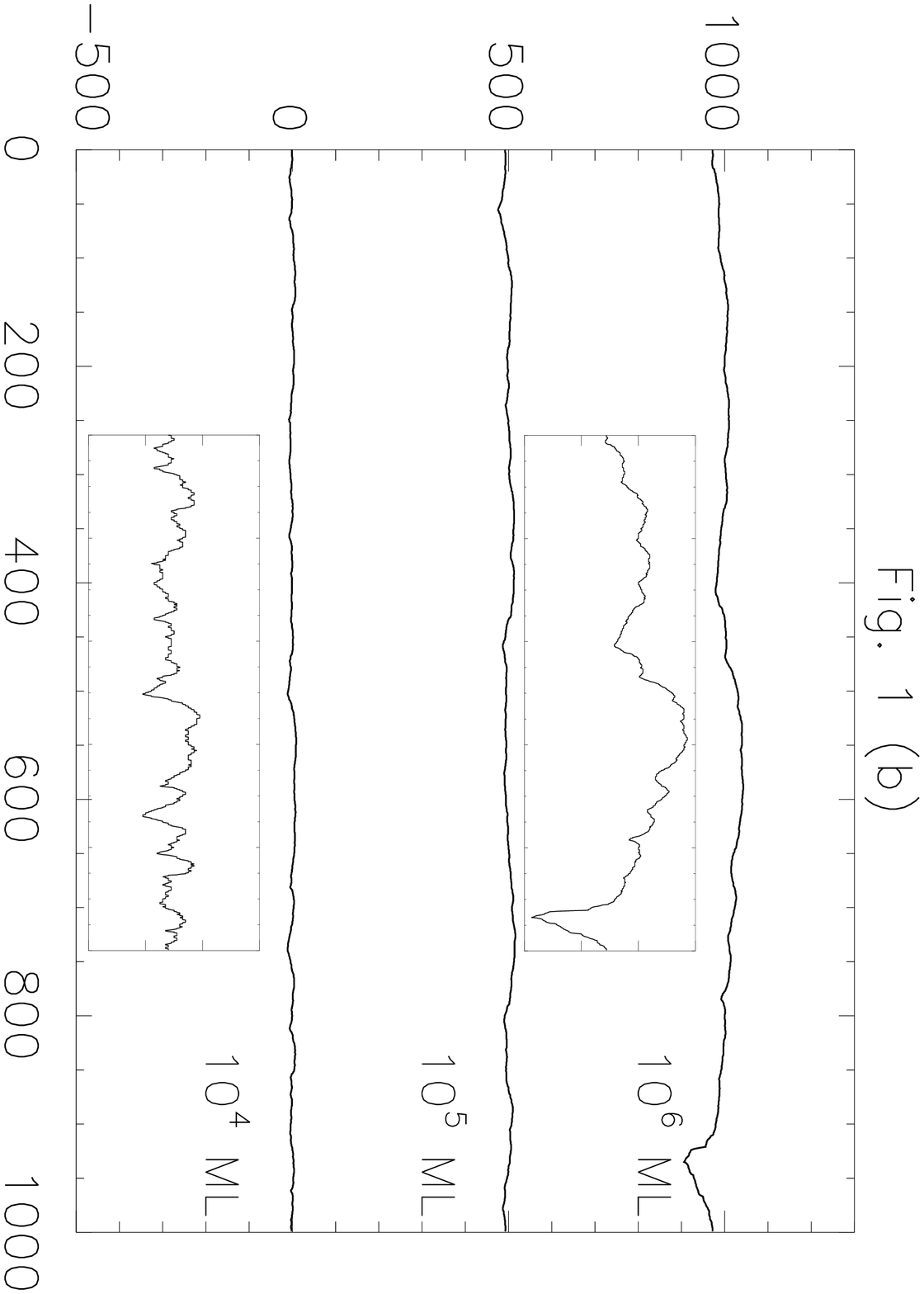}
   }
  \hss}
 }
\caption{
(a) Dynamical morphologies from the system of substrate size $L=1000$
at $10^4$, $10^5$ and $10^6$ monolayers (ML) for the original model,
ie. $m=1$.
Inset: Schematic configuration defining growth rules in 1+1 dimensions,
$m$ hits are needed at a site for an actual deposition event
($m=1$ in ref. 1).
(b) Morphologies from the $m=10$ model plotting on the exact same scale
as in (a) showing the much smoother surfaces. The small insets show
the morphologies at $10^4$ and $10^6$ ML in appropriated
expanded scales so that
the detailed rough morphology can be seen.
}
\end{figure}
(1) Recent large scale simulations \cite{12} of the correspending 
two dimensional (d=2+1) growth model seem to fairly unambiguously
indicate the 2+1 - dimensional growth universality class to be that
of the fourth-order nonlinear conserved MBE growth equation \cite{13};
(2) a number of theoretical approaches based on the kinetic master
equation technique \cite{14,15,16} as well as symmetry arguments 
\cite{9} lead to the conclusion that the one dimensional model 
should belong to the fourth-order nonlinear conserved MBE growth
equation; (3) extensive large scale simulations (using upto $10^{14}$
deposited atoms in the largest simulations) in d=1+1 produce
\cite{4,5,6,7,8,9} {\it excellent} scaling of the dynamically
evolving surface roughness with the scaling exponents, however,
being approximately consistent with the {\it linear} fourth-order
conserved growth equation \cite{13} rather than the expected 
nonlinear one (with the additional complication \cite{7,8} of 
there being substantial skewness in the growth morphology
implying that a {\it linear} description cannot apply). 
In addition, the model exhibits an intriguing anomalous
multiscaling \cite{6,7,8,9,10,11,12} behavior in the height
correlation functions, which transcends the standard 
self-affine dynamic scaling ansatz.
In this paper we obtain the elusive ``correct'' 
asymptotic universality class
of this one dimensional \cite{1} discrete growth model by
introducing a multiple hit noise reduction scheme 
which has earlier been
successful \cite{17,18} in the identification of the growth
universality classes for Eden and ballistic deposition models.
Our noise reduction results establish 
that the discrete limited mobility growth model introduced
in ref.1 does indeed belong to the fourth-order {\it nonlinear}
conserved MBE growth equation.

\begin{figure}

 \vbox to 5.5cm {\vss\hbox to 6cm
 {\hss\
   {\includegraphics{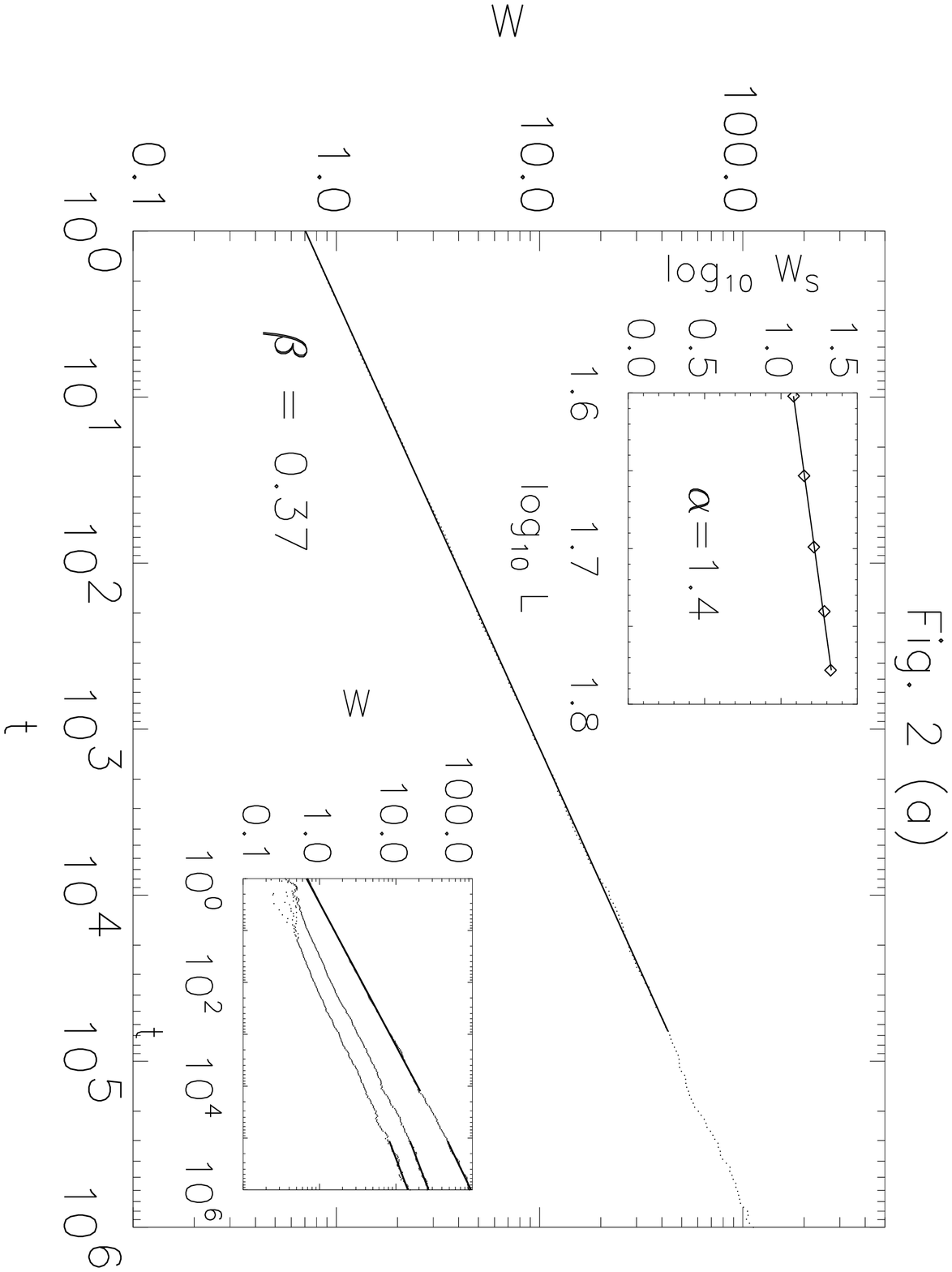}
   }
  \hss}
 }

\vskip .5cm

 \vbox to 5.5cm {\vss\hbox to 6cm
 {\hss\
   {\includegraphics{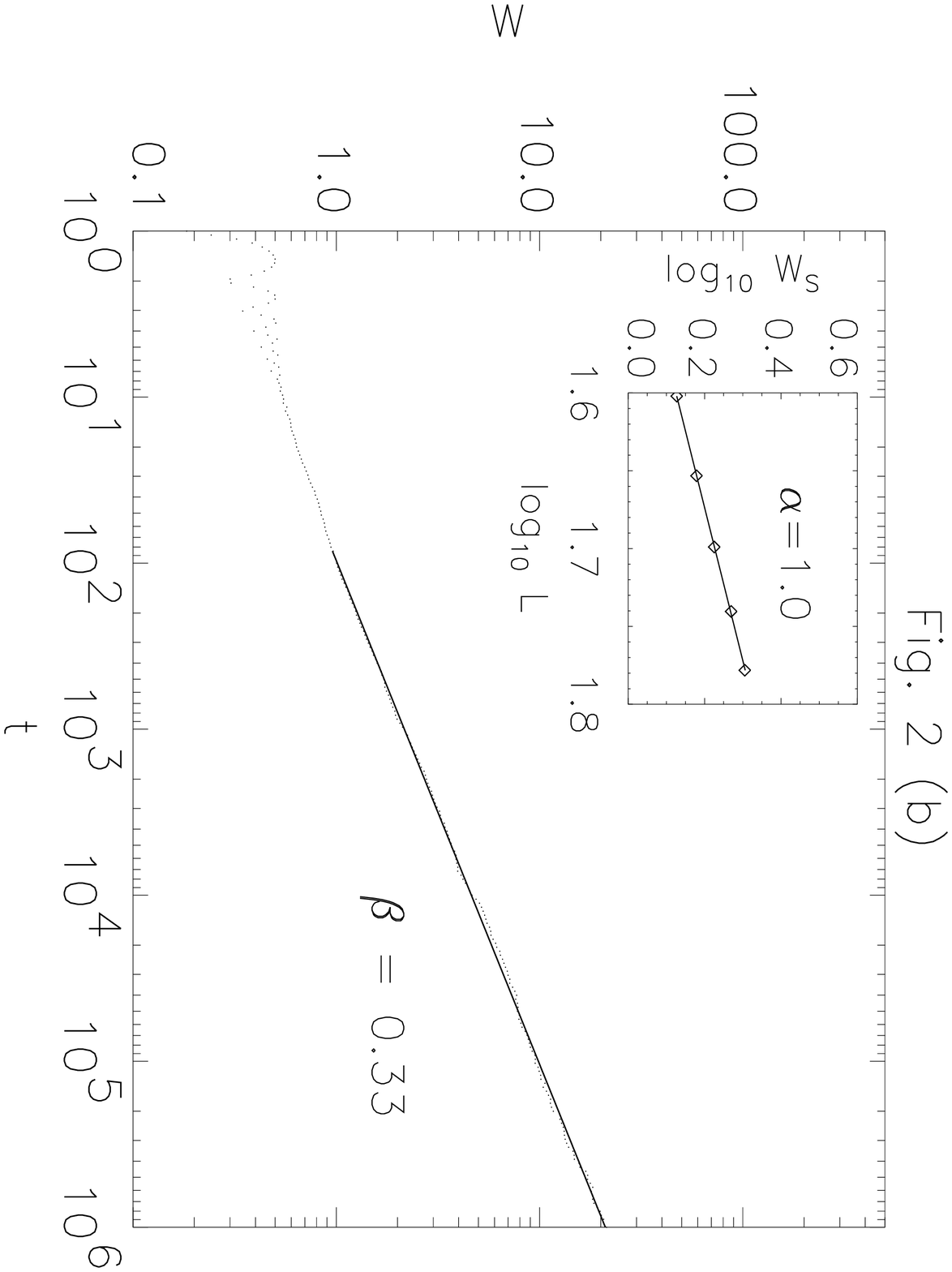}
   }
  \hss}
 }
\caption{
The interface width $W(t)$ as a function of deposition time. Solid
lines indicate the power law fit with the growth exponent $\beta$
for system $L=1000$ with (a) $m=1$ and (b) $m=10$.
Inset: Plots of $\log_{10} W_s$ vs $\log_{10} L$ where $W_s$ is
the saturation width. Slopes yield the roughness exponent $\alpha$.
Right inset in Fig. 2 (a): $W(t)$ for WV model with
$m$ = 1, 5, and 15 from top to bottom. In the original model
($m=1$) $\beta \approx 0.36$ whereas in the noise reduced
($m$ = 5, 15) results $\beta \approx 0.26$.
}
\end{figure}
Our growth model \cite{1} is shown in Fig. 1, where we also show the
dynamical growth morphologies for the original model and the
noise reduced model. The noise reduction scheme rescales time
by the noise reduction factor $m$ (where $m$ is the number of
attemps required at a site for an actual deposition process
to occur --- $m=1$ in the original model of ref.1), and all our
times are given in terms of this rescaled time (which also defines
the average film thickness in our model). The most striking 
aspect of Fig. 1 is that noise reduction is seen to have a drastic
effect on the growth morphology --- it strongly reduces the 
high surface steps and deep grooves which were the hallmark of the
original \cite{1} limited mobility growth model. This suppression 
of high steps (equivalently, the reduction of local slopes) in
the growing surface is, in fact, the key to the success of the 
noise reduction scheme in obtaining the asymptotically correct 
universality class of the growth model. We emphasize, however, 
that noise reduction, while drastically suppressing surface
high steps/deep grooves, still produces a skewed growth morphology
where the up-down symmetry of the starting flat substrate is
spontaneously broken by the nonequilibrium growth process
--- in fact the measured skewness in the growth morphology
is found to be independent of the 
noise reduction scheme. The skewness ($\approx -0.5$) in the
growth morphology (even in its saturated steady state) is a
unique characteristic of the nonequilibrium growth model of
ref.1 which is not shared by the other limited mobility epitaxial 
growth models \cite{21,22} existing in the literature.

To proceed further we use the dynamical scaling ansatz, and
discuss the evolving surface kinetic roughness in terms of 
two independent critical exponents $\beta$ and $\alpha$.
The interface width or the root mean square fluctuation in
the dynamical surface height $h(x,t)$, where $h$ is the height
of the growing surface at (reduced) time $t$ at the substrate
spatial point $x$, is defined as 
$W(L,t) \equiv \langle (h-\langle h \rangle )^2 \rangle ^{1/2}$
with the angular brackets representing average over the 
substrate of width $L$ (in the one dimensional $x$ direction)
as well as a noise ensemble average. The dynamic scaling ansatz
(which is obeyed extremely well by our results for all values
of the noise reduction factor $m=1$ -- $15$ we investigated)
asserts that $W(L,t) \sim t^{\beta}$ for $\xi \ll L$ and 
$W(L,t \rightarrow \infty) \equiv W_s(L) \sim L^{\alpha}$
in the saturated steady state for $\xi \gg L$, where the lateral
correlation length $\xi \equiv \xi(t) \sim t^{1/z}$.
The growth(roughness) exponent(s) $\beta (\alpha)$ and the dynamical
exponent $z = \alpha / \beta$ (which describes the approach 
to the steady state associated with the growth of lateral
correlations) completely define the universality class of
the growth model provided the growth problem is self-affine.
In Fig. 2 we show our calculated $W(t)$ and $W_s(L)$ dynamic
scaling plots for various values of the noise reduction factor.
While all the results show excellent dynamic scaling, it is 
clear that the critical exponents for the noise reduced model
are $\beta \approx 0.33$, $\alpha \approx 1.0$, $z \approx 3$ whereas
the original model ($m=1$) gives $\beta \approx 0.37$, $\alpha \approx
1.4$, $z \approx 3.9$ in agreement with earlier findings 
\cite{1,4,7,8,9,10,11}. We believe that the critical exponents
$\beta \approx 0.33$, $\alpha \approx 1.0$, $z \approx 3$ are 
the correct asymptotic exponents defining the universality class 
of the one dimensional growth model originally introduced in 
ref.1, and that the noise reduction scheme successfully suppresses 
the correction to scaling which dominates the $m=1$ version
of the model for many decades in time.

The coarse grained continuum equation which is believed
\cite{7,8,9,10,11} to describe the growth model of ref.1 is the
conserved nonlinear MBE growth equation \cite{13} :
\begin{equation}
\frac{\partial h}{\partial t} = - \nu_4 \frac{\partial^4 h}{\partial x^4}
+ \lambda_4 \frac{\partial^2}{\partial x^2} \left( \frac{\partial h}
{\partial x} \right) ^2 + \sum_{n=3}^{\infty} \lambda_{2n} 
\frac{\partial^2}{\partial x^2} \left( \frac{\partial h}{\partial x}
\right) ^{2n} + \eta ,
\label{eqn1}
\end{equation}
where $h \equiv h(x,t)$ is now the height fluctuation, 
$h - \langle h \rangle$, around the average surface height,
and $\eta(x,t)$ is the (white) shot noise associated with the
deposition beam fluctuations which produce the kinetic surface
roughening. The critical exponents for the corresponding fourth
order {\it linear} equation \cite{1,22}, where 
$\lambda_4 = \lambda_{2n} = 0$, are trivially known to be
$\beta = 0.375$, $\alpha = 1.5$, $z=4$. 
It was already noted in ref.1 that the simulated growth exponents
of the discrete model seem to be following those of the linear
version of Eq.(\ref{eqn1}) although no particular significance
was attached to this fact. The intriguing aspect of the critical
behavior of the discrete model in d=1+1 dimensions has been 
its consistency with the linear version of Eq.(\ref{eqn1})
as far as the {\it global} exponents, $\beta$ ($\simeq 0.375$),
$\alpha$ ($\simeq 1.5$), and $z$ ($\simeq 4$) go 
whereas at the same time the
up-down symmetry of the growth problem, which is manifestly
present in the linear equation because $h \rightarrow -h$
leaves the equation invariant, is broken with the evolving
growth morphology explicitly showing a finite skewness
$s = \langle h^3 \rangle \langle h^2 \rangle ^{-3/2}
\simeq -0.5$ (obviously $s \equiv 0$ for the linear equation).
Thus the puzzle until our current work has been that the
one dimensional discrete growth model should {\it not} 
belong to the fourth order conserved linear growth equation
universality, except that it does for as long as (at least upto
eight decades in time) one can dynamically simulate the model
\cite{1,2,3,4,5,6,7,8,9,10,11,12}.
Our multiple hit noise reduction scheme resolves the mystery
by obtaining the
asymptotic exponents by successfully eliminating the severe
correction to scaling problem which hinders the $m=1$ version
\cite{1} of the model. This is similar to what was earlier 
found in the Eden model \cite{17,18}.

The critical exponents we obtain in the noise reduced model 
($\beta \simeq 0.33$, $\alpha \simeq 1.0$, $z \simeq 3$) are
consistent with the 1-loop dynamical RG \cite{13} treatment
($\beta = 1/3$, $\alpha = 1$, $z=3$) and direct numerical 
simulations \cite{11,23} of the {\it fourth order nonlinear
conserved MBE} growth equation \cite{13}
with $\lambda_{2n}=0$, but $\nu_4 , \lambda_4
\neq 0$ in Eq.(\ref{eqn1}). The original thinking \cite{13}
that these one loop results may in fact be exact for the 
fourth order nonlinear equation has recently been questioned
\cite{24} with a 2-loop dynamical RG treatment obtaining 
miniscule (less than $0.5\%$) numerical corrections. 
These 2-loop corrections are too small to be of any practical
significance to our work (or to other simulations and 
experiments).
We therefore conclude that the critical exponents of the 
noise reduced version of the growth model introduced in 
ref.1 are the same as those of Eq.(\ref{eqn1}) with
$\lambda_{2n} \equiv 0$ for $n \geq 3$, and therefore the model
belongs to the universality class of the nonlinear fourth
order conserved MBE growth equation.

We believe that the success of noise reduction in producing
the correct asymptotic growth universality arises from the suppression
of the infinite series of nonlinear term ($\lambda_{2n} \neq 0$)
in Eq.(\ref{eqn1}) which are associated with the high steps in the
original ($m=1$) discrete growth model \cite{1}.
The role of the infinite series of nonlinear terms (i.e.
$\lambda_{2n} \neq 0$ for $n=3,4,...$) in Eq.(\ref{eqn1})
is quite subtle \cite{10,11} because the
existence of such an infinite series of {\it relevant}
terms is unusual in 
critical phenomena. 
High steps, however, imply large values of the slope
$\partial h / \partial x$, indicating the existence of 
the infinite series.
Simple power counting shows all
the terms in this infinite series to be {\it marginally
relevant operators} in the one dimensional growth problem,
and since they are all allowed by the symmetry of the discrete
model, they should all be present \cite{7,11} in the growth
model of ref.1. 
Note that the corresponding linear problem has 
$\alpha$ ($= 1.5$) $> 1$,
implying that the infinite nonlinear series should be 
generated if allowed by symmetry.
This infinite series has recently been shown 
\cite{10,11} to be responsible for giving rise to the 
anomalous ``intermittent'' multiscaling behavior in the growth
model of ref.1 which has attracted 
considerable attention \cite{6,7,8,9,10,11}.
In Fig. 3 we show that the noise reduction 
\begin{figure}

 \vbox to 5.5cm {\vss\hbox to 6cm
 {\hss\
   {\includegraphics{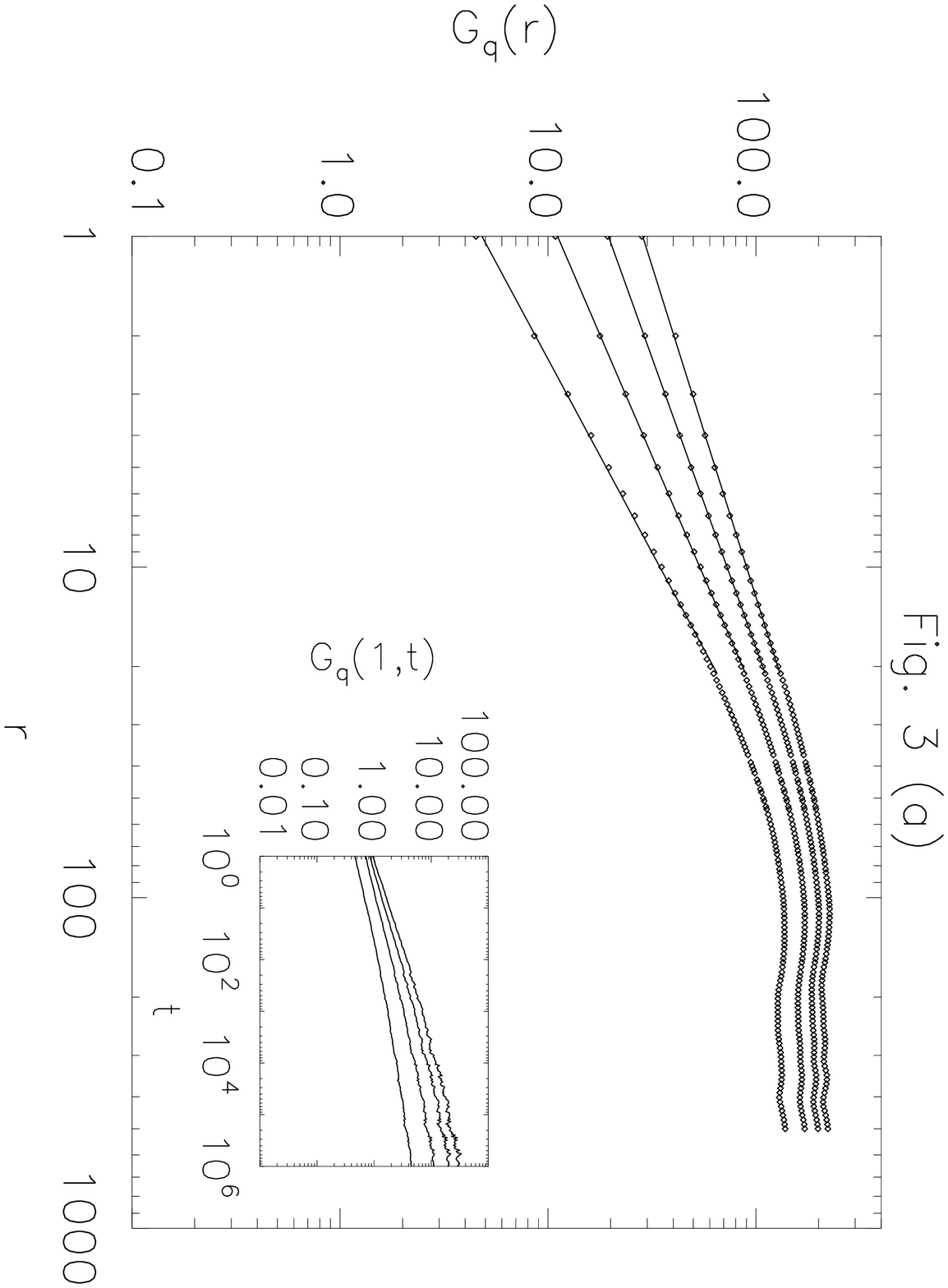}
   }
  \hss}
 }

\vskip .5cm

 \vbox to 5.5cm {\vss\hbox to 6cm
 {\hss\
   {\includegraphics{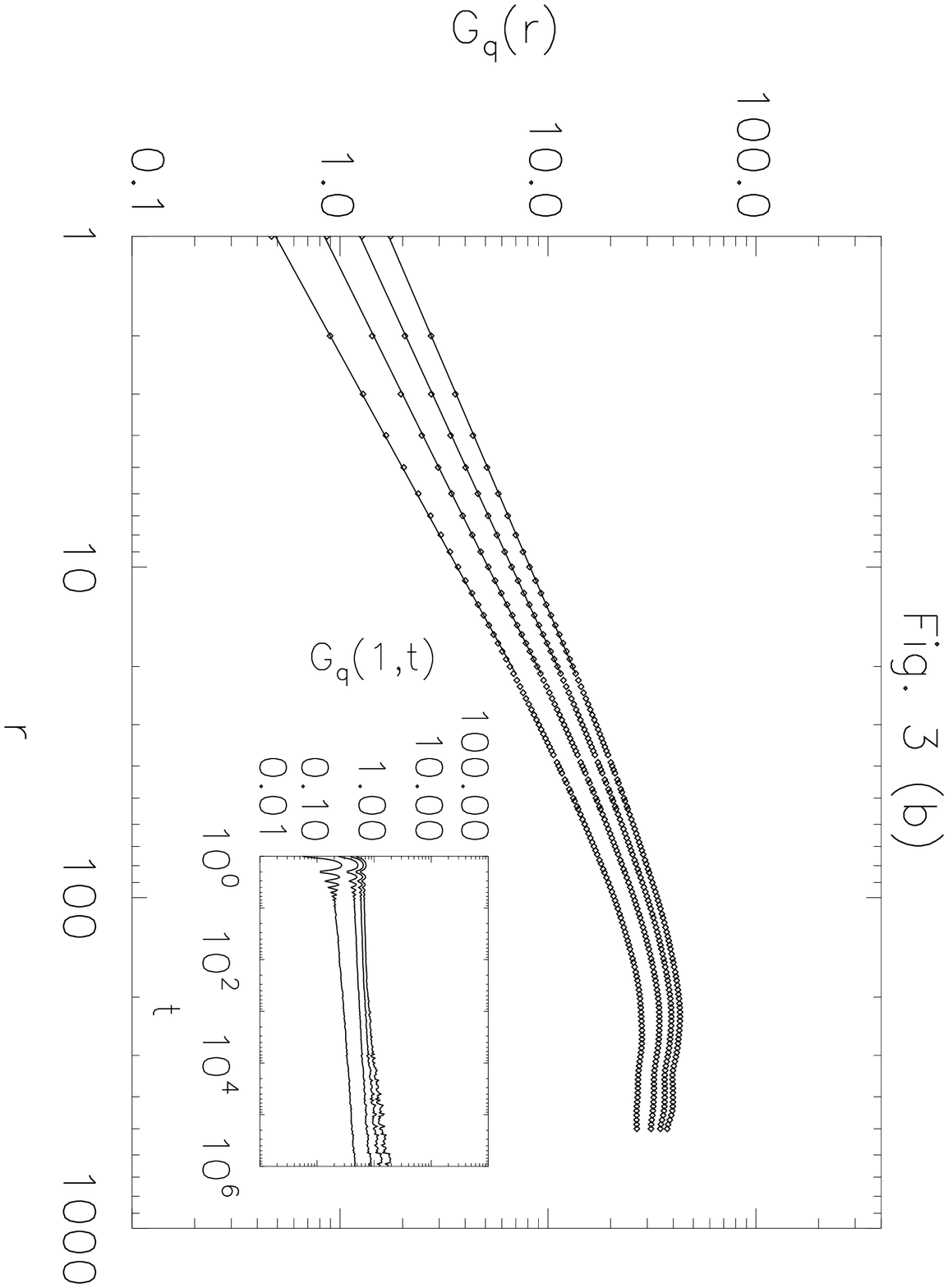}
   }
  \hss}
 }
\caption{
(a) Anomalous multiscaling behavior of the height correlation functions
$G_q(r,t) = \langle |h(x+r,t)-h(x,t)| ^q \rangle ^{1/q}$
at fixed $t=10^6$ ML and the nearest-neighbor height difference
correlation functions $G_q(1,t)=\langle |h(x+1,t)-h(x,t)| ^q \rangle ^{1/q}$
(inset) with $q=1-4$ from bottom
to top. Substrate size $L=1000$ and $m=1$.
(b) The same plots for system with $m=10$.
The noise reduced correlation functions show only very
weak anomalous multiscaling behavior.
}
\end{figure}
technique essentially eliminates the anomalous multiscaling
behavior in the height correlation functions of the 
growth model of ref.1. Thus the noise reduction
technique resolves all three of the intriguing and puzzling
features of the one dimensional minimal MBE growth model
of ref.1,
namely, (1) it establishes beyond any reasonable doubt 
the correct universality class of the model to be the
`expected' fourth order conserved nonlinear continuum growth
equation; (2) it suppresses the ``unphysical'' feature of
high steps and deep grooves in the growth morphology
while maintaining a finite skewness in the morphology;
(3) it suppresses the anomalous multiscaling in the model.

Before concluding we emphasize an important feature of our 
growth model which is characterized by what is {\it absent}
from Eq.(\ref{eqn1}) --- the well-known \cite{9,25,26}
Laplacian $\nu_2 \frac{\partial^2 h}{\partial x^2}$ term
associated with the generic Edwards - Wilkinson (EW) universality
class is strictly absent from the growth equation describing
the discrete growth model of ref.1. The model of ref.1 is,
in fact, the only known limited mobility MBE growth model
{\it which does not belong to the generic EW growth
universality class}. It is worth emphasizing this point 
because this has been a controversial and contentious
\cite{2,3,9,27} issue in the past. 

To reinforce this point we have also carried out noise reduction
simulations of the closely related Wolf-Villain (WV) model
\cite{22}, which differs from the model of ref.1 only in that
all deposited atoms, {\it independent of their initial
coordinations}, are allowed to move to lateral nearest-neighbor
sites in order to {\it maximize} their local coordination number.
Although it is well-accepted \cite{9,27} that the WV model 
asymptotically belongs to the EW universality class, this
crossover has never been clearly observed in simulations
because for all practical purposes the dynamic scaling behavior 
of the WV model \cite{22} is similar to that of the model of 
ref.1 upto the longest simulation times. Our d=1+1 noise reduced
WV model simulations, shown as an inset in Fig. 2 (a),
clearly show that the asymptotic growth exponent $\beta$
decreases from $\sim 0.36$ (for $m=1$) to $\sim 0.26 \simeq
\beta_{EW}$ in d=1+1 (for $m=15$) under the noise reduction
technique. Thus the WV model belongs to the EW universality
and the model of ref.1 belongs to the fourth order nonlinear
MBE growth universality.

The non-existence
of the EW term (which, if it existed, would have defined
the universality class of the model because all the fourth
order terms in Eq.(\ref{eqn1}) are {\it irrelevant} compared
with the EW Laplacian term) in the growth model of ref.1 is,
in fact, an exact result due to a hidden symmetry \cite{9}
in the growth model which produces an exactly vanishing surface
current \cite{27} in the model on a tilted substrate.
The surface current on a tilted substrate is, in general,
proportional \cite{9,27} to the strength $\nu_2$ of the
generic EW term in the growth equation, and therefore its
vanishing implies $\nu_2 \equiv 0$. 
Our calculated inclination dependent surface current on a tilted
substrate in the growth model is essentially zero within 
error bars 
for all values of the noise reduction factor ($m$ = 1 -- 15).
This is also a strong indication that noise reduction does not
change the universality class of our growth model.  
This is similar to what was found
earlier \cite{17,18} in the Eden model. It may be worthwhile 
to point out that in the Eden model also \cite{17,18}
the success of noise reduction in eliminating corrections
to scaling arises from the suppression of high steps.

We conclude by discussing why understanding the 
universality class of the growth model introduced in ref.1
is of considerable interest. An important theoretical reason
is that this model is the {\it only} limited mobility MBE
growth model which does not belong to the generic EW universality
class, and therefore, as an exception, its proper theoretical
understanding is of obvious interest. The fact that this 
growth model exhibits complex and highly nontrivial anomalous
multiaffine dynamic scaling \cite{6,7,8,9,10,11} is an
additional theoretical incentive in understanding its growth
universality. Another significant feature is that, by 
construction, this growth model is the low temperature
version \cite{1,4} of the full temperature dependent
activated diffusion MBE growth model \cite{4,8} 
because in the limited mobility model only the adatoms
without any lateral bonding are allowed to increase 
their coordination through diffusion,
and therefore it has considerable experimental significance.
It may be appropriate in this context to point out that
several experimental measurements 
\cite{28} of MBE growth exponents
($\beta \approx 0.2$, $\alpha \approx 0.7$) are consistent 
with the (d=2+1 dimensional) critical exponents given by
the fourth order nonlinear conserved growth equation which,
as we show in this paper 
and in ref.12, defines the universality class of the
limited mobility discrete growth model of ref.1.

This work has been supported by the US-ONR and the
NSF-MRSEC.

\end{document}